\documentclass{article} 
\usepackage{iclr2023_conference,times}

\usepackage{amsmath,amsfonts,bm}

\def\eqref#1{equation~\ref{#1}}

\def\1{\bm{1}}

\DeclareMathAlphabet{\mathsfit}{\encodingdefault}{\sfdefault}{m}{sl}
\SetMathAlphabet{\mathsfit}{bold}{\encodingdefault}{\sfdefault}{bx}{n}

\usepackage{hyperref}
\usepackage{url}
\usepackage{graphicx}
\usepackage{booktabs}
\usepackage{wrapfig}
\usepackage[textsize=tiny]{todonotes}

\newcommand{\ours}{\textsc{DiffDock-PP}}

\title{DiffDock-PP: Rigid Protein-Protein Docking with Diffusion Models}

\author{Mohamed Amine Ketata$^{*1}$, Cedrik Laue$^{*1}$, Ruslan Mammadov$^{*1}$, Hannes Stärk$^2$, \\ \textbf{Menghua Wu$^2$, Gabriele Corso$^2$, Céline Marquet$^1$, Regina Barzilay$^2$, Tommi S. Jaakkola$^2$}
\\
\\
$^1$Technical University of Munich, Germany \;
$^2$Massachusetts Institute of Technology, USA\\
\texttt{\{mohamedamine.ketata,cedrik.laue,ruslan.mammadov\}@tum.de} 
}

\iclrfinalcopy 
\begin{document}

\def\thefootnote{*}\footnotetext{Equal contribution}
\def\thefootnote{\arabic{footnote}}

\maketitle

\begin{abstract}
Understanding how proteins structurally interact is crucial to modern biology, with applications in drug discovery and protein design. Recent machine learning methods have formulated protein-small molecule docking as a generative problem with significant performance boosts over both traditional and deep learning baselines. In this work, we propose a similar approach for rigid protein-protein docking: \textsc{DiffDock-PP} is a diffusion generative model that learns to translate and rotate unbound protein structures into their bound conformations. We achieve state-of-the-art performance on DIPS with a median C-RMSD of 4.85, outperforming all considered baselines. Additionally, \textsc{DiffDock-PP} is faster than all search-based methods and generates reliable confidence estimates for its predictions.\footnote{Our code is publicly available at \texttt{https://github.com/ketatam/DiffDock-PP}}
\end{abstract}


\section{Introduction}
Proteins realize their myriad biological functions through interactions with biomolecules, such as other proteins, nucleic acids, or small molecules.
The presence or absence of such interactions is dictated in part by the geometric and chemical complementarity of participating bodies.
Thus, learning how individual proteins form complexes is crucial to understanding protein activity.
In this work, we focus on rigid protein-protein docking: given two protein structures, the goal is to predict their resultant complex while maintaining internal bonds, angles, and torsion angles fixed.

Traditional approaches for rigid protein-protein docking consist of a search algorithm followed by a scoring function \citep{chen2003zdock,de2010haddock,yan2020hdock}. After enumerating a vast search space of potential poses, these methods rely on heuristics or empirical methods to select the most plausible poses. Due to the exhaustive search required, these methods are often slow and computationally expensive. More recently, deep learning approaches have tackled protein-protein docking as a regression problem: given two structures, directly predict the final pose \citep{ganea2021independent,ppi}. While fast, these models have yet to outperform search-based algorithms.

\begin{figure}
\centering
\vspace{-1cm}
\includegraphics[width=\textwidth]
{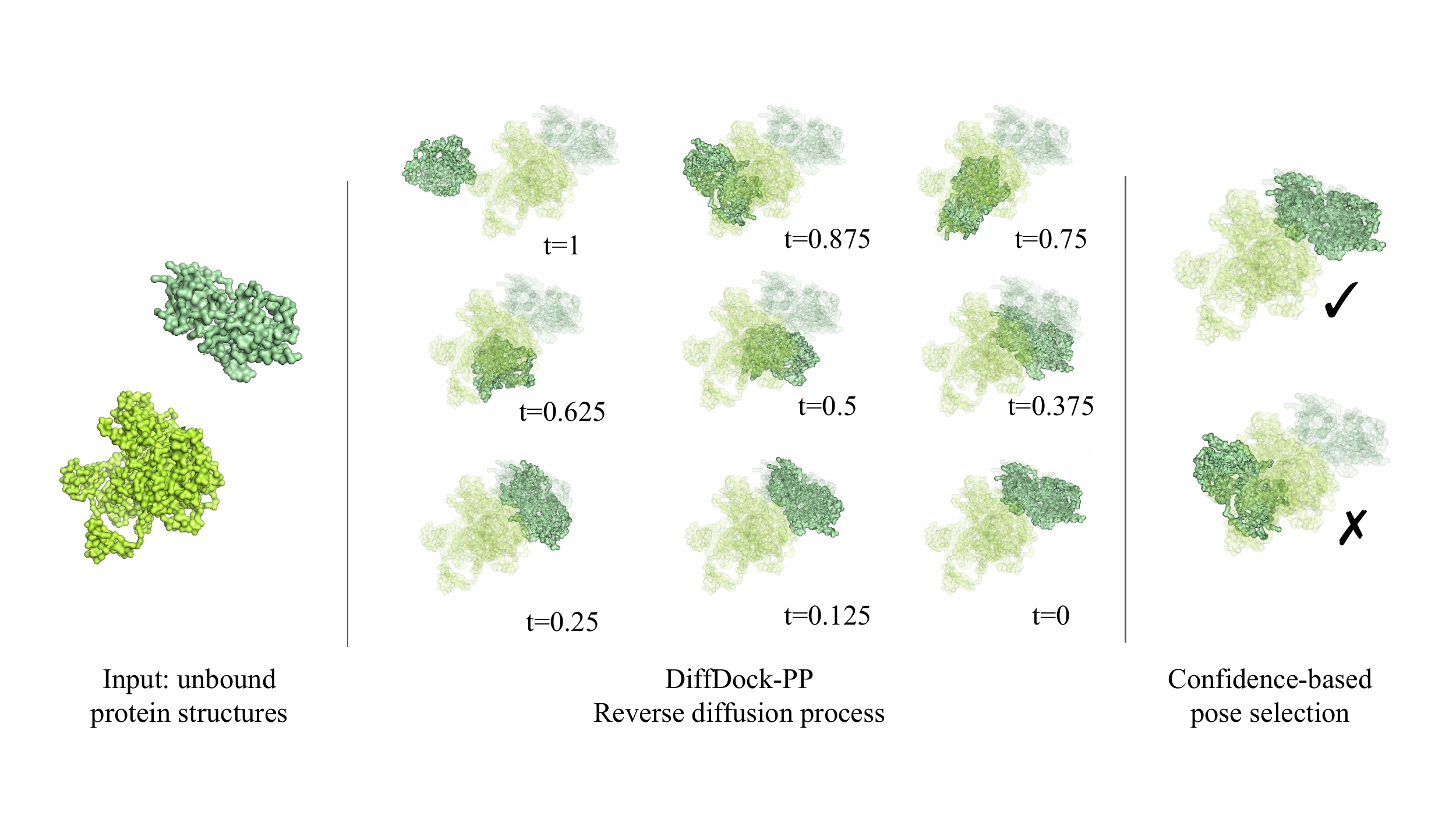}
\vspace{-1cm}
\caption{Overview of \ours{}. The model takes two proteins as input, where the ligand has been randomly rotated and translated in 3D space. Then it runs a reverse diffusion process to sample multiple poses. The confidence model ranks these poses, and we output the pose with the highest confidence score. The ground truth pose is shown in light grey and is added to all diffusion steps. Depicted structure is PDB 1KEE over 40 steps of reverse diffusion.}\label{fig:overview}
\end{figure}

Inspired by recent breakthroughs in protein-small molecule docking \citep{corso2022diffdock}, we instead propose that protein-protein docking be formulated as a generative problem: given two proteins, the goal is to estimate the distribution over all potential poses using a diffusion generative model. To obtain the final docked pose, we sample from this distribution multiple times and select the best one via a learned confidence model, as shown in Figure \ref{fig:overview}. We call our method \ours{}.


Empirically, \ours{} achieves a top-1 median complex root mean square deviation (C-RMSD) of 4.85 on the Database of Interacting Protein Structures (DIPS), outperforming all considered baselines. Compared to popular search-based docking software, \ours{} is 5 to 60 times faster on GPU.

\section{Background and Related Work}
\label{background}

\textbf{Protein-Protein Docking.} The goal of protein-protein docking is to predict the (bound) structure of a protein-protein complex based on the individual proteins' (unbound) structures. Specifically, we focus on the task of \textit{rigid body} protein-protein docking, which assumes that the proteins do not undergo any deformations during binding, restricting their relative degrees of freedom to a rotation and translation in 3D space. This assumption is often realistic \citep{pp-docking} and even leads to improved results for most interacting proteins \citep{desta2020performance}.

To evaluate the quality of the predicted structures, a common approach is to compute the fraction of those lying within some threshold distance to the true bound structure \citep{pp-docking,dockq,capri}.



\textbf{Search-based Docking Methods.}
Traditional methods for protein-protein docking usually rely on the physical properties of the complexes \citep{chen2003zdock,de2010haddock,yan2020hdock}. These methods typically 1) generate an initial population of plausible complex structures, 2) further pose proposals using optimization algorithms, and 3) refine the complexes with the highest score according to some scoring function.
Template-based modeling (TBM), which predicts the structure of a target protein by aligning it to one or multiple template proteins with known structures, is also used as a subroutine by some search-based methods \citep{pp-docking}.
While some of these methods offer decent predictive performance, they are usually computationally expensive and impractical for large-scale molecular screening campaigns.


\textbf{Deep Learning-based Docking Methods.}
Deep learning approaches to protein-protein docking can be broadly partitioned into two categories: single-step and multi-step methods. Single-step methods directly predict the complex structure in a one-shot fashion. Notably, \citet{ganea2021independent} proposed \textsc{EquiDock}, a pairwise-independent SE(3)-equivariant graph matching network that directly predicts the relative rigid-body transformation of one of the interacting proteins. Furthermore, \citet{sverrisson2022physicsinformed} incorporated different physical priors into an energy-based model for predicting protein complex 3D structure.
In contrast, multi-step methods produce their final predictions by iteratively refining a set of proposed structures.
For instance, \textsc{AlphaFold-multimer} \citep{alphafold_multimer} was designed to co-fold multiple protein structures, given their primary sequences and multiple sequence alignments (MSAs) to evolutionary-related proteins.
In parallel with this work, \cite{mcpartlon2023deep} proposed \textsc{DockGPT}, a generative protein transformer for flexible and site-specific protein docking.


Our method, \ours{}, naturally falls into the category of multi-step methods due to the multiple steps required to sample from the distribution induced by diffusion generative models. Compared to search-based methods, however, we sample orders of magnitude fewer poses during our refinement process.



\textbf{Diffusion Generative Models (DGMs).} DGMs offer a powerful way to represent probability distributions beyond likelihood-based and implicit generative models, circumventing many of their issues. The main idea is to define a diffusion process transforming the data distribution in a tractable prior and learn the \emph{score function}, which is the \emph{gradient} of the log probability density function $\nabla_\mathbf{x}\log p_t(\mathbf{x})$\footnote{Note that this is a very different concept from the \textit{scoring function} of search-based docking methods.}, of this evolving distribution.
We can then use the learned score function to sample from the underlying probability distribution using well-established algorithms \citep{diffusion}.
A plethora of DGMs have been developed for tasks in computational biology, including conformer generation \citep{jing2022torsional}, molecule generation \citep{hoogeboom2022equivariant}, and protein design \citep{trippe2022diffusion}. The foundation of our method is \textsc{DiffDock} \citep{corso2022diffdock}, a diffusion generative model over the product space of the ligand's degrees of freedom (translational, rotational, and torsional). We extend this approach to the protein docking task.

\section{Method}
\subsection{Benefits of Generative Modeling for Rigid Protein Docking}
Protein-protein docking is often evaluated on the basis of thresholding \citep{dockq,capri}, e.g., a Ligand-RMSD $< 5$ \AA{} and an Interface-RMSD $< 2$ \AA{} are among several conditions to consider a given prediction to be of medium quality in \citet{capri}. 


Following the arguments of \citet{corso2022diffdock} and noting that directly optimizing such thresholding-based objectives is not feasible because they are not differentiable, we argue that these objectives are better aligned with training a \textit{generative} model to maximize the likelihood of the observed structures than with fitting a \textit{regression} model as done in previous work. Concretely, since real-world data, as well as complex deep learning models, suffer from inherent multi-modal uncertainty, regression-based methods trained to predict a single pose that, in expectation, minimizes some MSE-type loss \citep{ganea2021independent} would learn to predict a structure as the weighted mean among many viable alternatives, often not a plausible structure itself. In contrast, a generative model would aim to capture the distribution over these alternatives resulting in more plausible, accurate, and diverse structures.

\begin{figure}
\centering
\includegraphics[width=0.7\textwidth]
{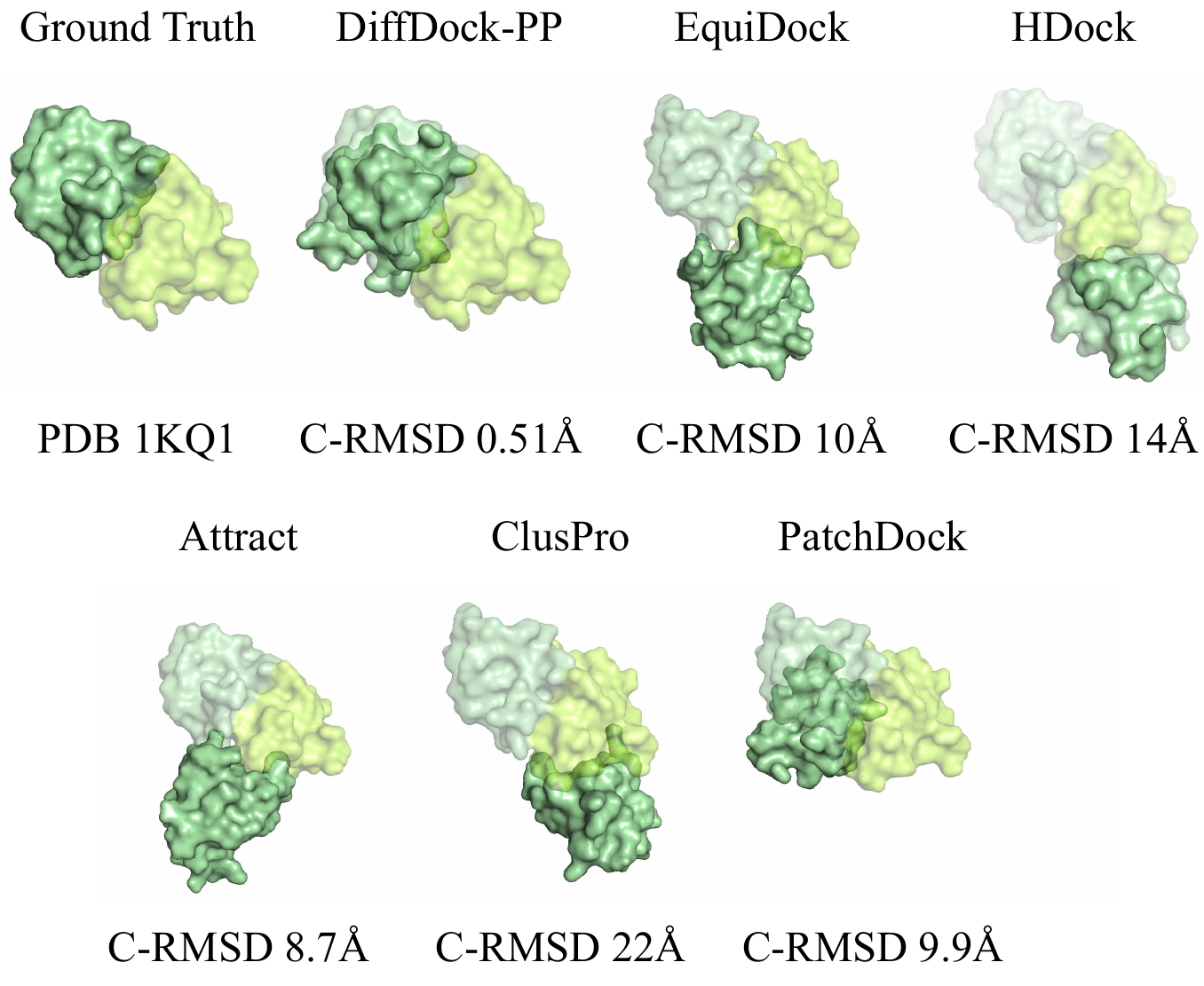}
\caption{Visualization of the different methods' predictions for PDB 1KQ1 from our test set. The ground truth pose is depicted in light grey.}\label{fig:predictions}
\end{figure}

To illustrate this phenomenon, we visualize some of the structures predicted by our model and compare them to those generated by the baselines, especially \textsc{EquiDock}, which was trained using an MSE-type loss and whose one of the main limitations is the existence of steric clashes in its predicted structures \citep{ganea2021independent}. Figure \ref{fig:predictions} illustrates such predictions. We observe that our model predicts structures with no steric clashes, which we hypothesize is in part due to the adopted generative approach to protein-protein docking.

\subsection{Method Overview}
As defined in Section \ref{background}, in rigid protein-protein docking, we aim to predict the complex structure of an interacting protein pair based on the individual structure of each protein. 

In this work, we model the proteins on the residue level, representing each protein as a set of amino acid nodes. Each residue is, in turn, represented by its type and the position of its $\alpha$-carbon atom. We denote $\mathbf{X}_1 \in \mathbb{R}^{3n}$ as the ligand consisting of $n$ residues and $\mathbf{X}_2 \in \mathbb{R}^{3m}$ as the receptor with $m$ residues. The ligand/receptor assignment can, in principle, be arbitrary; however, we define the ligand $\mathbf{X}_1$ as the protein with fewer residues. This means that with $\mathbf{X}_1^* \in \mathbb{R}^{3n}$ and $\mathbf{X}_2^* \in \mathbb{R}^{3m}$ denoting the ground truth complex, the receptor is kept fixed $\mathbf{X}_2 = \mathbf{X}_2^*$ and the task is to predict the structure of the ligand with respect to the receptor.

It is important to note that - since we are considering \emph{rigid-body} protein docking - we only need to consider poses that can be obtained by a \emph{rigid-body} transformation (i.e., a rotation and a translation) of the initial pose $\mathbf{X}_1$. These poses lie on a 6-dimensional submanifold $\mathcal{M} \subset \mathbb{R}^{3n}$ corresponding to the 6 degrees of freedom introduced by the rotation and translation in 3D space. We consider rigid-body protein-protein docking as the task of learning a probability distribution $p(\mathbf{X}_1 | \mathbf{X}_2)$ of ligand poses in $\mathcal{M}$, conditioned on the receptor structure $\mathbf{X}_2$.

Now we discuss how we can effectively deploy DGMs to learn this probability distribution. In order to avoid the inefficiencies arising from learning DGMs on arbitrary submanifolds \citep{de2022riemannian}, we use the framework of intrinsic diffusion models \cite{corso2023modeling}: map the extrinsic submanifold $\mathcal{M}$ to the intrinsic manifold defined by the product space of the rotation and translation group and define our DGM over this manifold.

\subsection{Diffusion Process}


To formalize the discussion in the previous section, let us introduce the 3D translation group $\mathbb{T}(3)$ and the 3D rotation group $SO(3)$ as well as their product space $\mathbb{P} = \mathbb{T}(3) \times SO(3)$. With this, we can define a rigid-body transformation as the mapping $A:\mathbb{P}\times \mathbb{R}^{3n} \rightarrow \mathbb{R}^{3n}$ with 
\begin{equation}
A((\boldsymbol{r},R),\boldsymbol{x})_i = R\left(\boldsymbol{x}_i-\overline{\boldsymbol{x}}\right)+\overline{\boldsymbol{x}} + \boldsymbol{r},
\end{equation}
where $\boldsymbol{x}_i\in \mathbb{R}^3$ corresponds to the position of the $i$-th residue and $\overline{\boldsymbol{x}}$ represents the center of mass of the ligand protein. This equation simply describes a rotation around the center of mass followed by a translation.

 The submanifold of ligand poses introduced informally in the previous section can now be described using this transformation as $\mathcal{M}=\{A\left(\left(\boldsymbol{r}, R \right),\mathbf{X}_1 \right) \vert \left(\boldsymbol{r}, R \right) \in \mathbb{P}\}$. For similar arguments to \citet{corso2022diffdock}, the map $A\left(\cdot, \mathbf{X}_1\right):\mathbb{P}\rightarrow \mathcal{M}$ is a bijection, which guarantees the existence of the inverse map. We can therefore develop a diffusion process over the product space $\mathbb{P}$ to generate a distribution on the manifold. 

 Given that $\mathbb{P}$ is a product manifold, we can define a forward diffusion process independently on each manifold \citep{rodola2019functional} with the score as an element of the corresponding tangent space \citep{de2022riemannian}. A score model can then be trained with denoising score matching \citep{song2019generative}. In both groups, we define the forward SDE as $d\bold{x} = \sqrt{d\sigma^2\left(t\right)/dt}\;  d\bold{w}$, where $\sigma^2$ is $\sigma_\text{tr}^2$ for $\mathbb{T}(3)$ and $\sigma_\text{rot}^2$ for $SO(3)$ and $d\bold{w}$ denotes the corresponding Brownian motion. The reader is referred to \citet{corso2022diffdock} for a description of how we can sample from and compute the score of the diffusion kernel on each of these groups.



\subsection{Model Architecture}


Both the score and confidence models are based on $SE(3)$-equivariant convolutional networks \citep{thomas2018tensor, geiger2020euclidean} adapted from \textsc{DiffDock}'s architecture \citep{corso2022diffdock} to mainly account for: i.) the \textit{symmetry} of protein-protein pairs, and ii.) the \textit{rigidity} assumption of the proteins. Below we summarise the main components of the architecture.

\textbf{Input representation.} Protein structures are represented as heterogeneous geometric graphs with the amino acid residues as nodes. Node features comprise the residue's type, the positions of its $\alpha$-carbon atoms, as well as language model embeddings trained on protein sequences from ESM2 \citep{esm}. In order to construct the edges, we connect each node to its 20 nearest neighbors from the same protein (intra-edges), and we use a dynamic cutoff distance of (40 + 3 * $\sigma_\text{tr}$)\AA{}, where $\sigma_\text{tr}$ is the current standard deviation of the diffusion translational noise to connect the nodes from different proteins (cross-edges). The intuition behind using a dynamic cutoff distance is to increase the chances that each node interacts with potentially relevant nodes from the other protein even when the proteins are still far apart (at early diffusion steps) while having a lower computational cost than using a fixed higher cutoff distance (especially at later diffusion steps).

\textbf{Intermediate layers.} After an initial set of embedding layers to process the initial features, the diffusion time, and the edge lengths, we define a different set of convolutional layers for each edge type (intra-edges and cross-edges). However, in contrast to \cite{corso2022diffdock}, we use the same intra-edge layers for both proteins to account for symmetry.

\textbf{Output layers.} This is where the main difference between the score and confidence model lies. On the one hand, the score model applies a tensor-product convolution placed at the center of mass of the ligand to produce two $SE(3)$-equivariant 3-dimensional vectors as the translational and rotational scores (lying in the tangent space of the respective manifolds). On the other hand, the confidence model applies a fully connected layer on the mean-pooled scalar representations from the last convolution layer to produce the $SE(3)$-invariant confidence value.

\subsection{Training and Inference}
The training and inference regimes follow very closely \citet{corso2022diffdock}. We reiterate the most important points here.

\textbf{Diffusion model.} Even though the diffusion kernel and score matching objectives were defined on the product space $\mathbb{P}$, we follow the extrinsic-to-intrinsic framework \citep{jing2022torsional} and develop the training and inference procedures directly on ligand-receptor poses in 3D space. This allows the model to reason about physical interactions more easily and should lead to better generalization. Another interesting point to note is that each training example $(\mathbf{X}_1, \mathbf{X}_2)$ is the only available sample from the conditional distribution $p(\cdot | \mathbf{X}_2)$. This is unlike the standard generative modeling setting, where many samples are drawn from the same data distribution. Therefore, during training, we iterate over distinct conditional distributions with only one sample. 
During inference, in order to avoid the problem of overdispersed distributions typically observed with generative models, we use low-temperature sampling \citep{temperatur}, which allows the model to concentrate on modes with high likelihood.

\textbf{Confidence model.}
The confidence model is a simple classification network trained to predict whether the structures sampled from the score model are of "good" quality, which we define as the structure having an L-RMSD below a certain threshold. To this end, we collect the training data for the confidence model by sampling from the (trained) diffusion model multiple times for each training complex and computing the L-RMSD for the sampled complexes. The labels are then generated by simply comparing the L-RMSD values to the threshold. In our experiments, we set the threshold to be 5\AA{}. The confidence model is then trained with cross-entropy loss. 


\textbf{Combined Inference.}
During inference, we sample a set of candidate poses from the diffusion model. These samples are then ranked by the confidence model according to the predicted confidence value of whether each pose has an L-RSMD below 5\AA{}. The final prediction is the pose with the highest confidence score.




\section{Experimental setup}


We evaluate our model on the Database of Interacting Protein Structures (DIPS) \citep{dips}.
DIPS consists of 42,826 binary protein complexes.
We use the same protein-family-based dataset split proposed by \cite{ganea2021independent}. 

\ours{} has 1.62M parameters and was trained on the DIPS training set for 170 epochs.
Every 10 epochs, we run reverse diffusion on the DIPS validation set to compute L-RMSD values. The best model obtained with this procedure is finally tested on the DIPS test set.
During training and inference, the smaller protein is selected as a ligand, and we randomly rotate and translate the ligand in space before running our model.


We compare our method to search-based docking algorithms \textsc{ClusPro (Piper)} (FFT-based) \citep{desta2020performance, kozakov2017cluspro}, \textsc{Attract} (based on a coarse-grained force field) \citep{schindler2017protein, de2015web}, \textsc{PatchDock} (based on shape complementarity principles) \citep{mashiach2010integrated, schneidman2005patchdock}, and \textsc{HDock} (makes use of template-based modeling and ab initio free docking) \citep{yan2020hdock, yan2017hdock, huang2014knowledge, huang2008iterative}.
For these baselines, we cannot control their training and testing data. This implies that some of them might have used a part of our test set to train or validate their models. 
Thus, the reported performance of these methods might be overestimating the true performance.

Furthermore, we compare our method to deep learning baselines \textsc{AlphaFold-multimer} \citep{alphafold_multimer} and \textsc{EquiDock} \citep{ganea2021independent}. 
Details on how we evaluated these baselines can be found in Appendix \ref{appendixA}.


To ensure a fair comparison, we follow the evaluation scheme proposed by \cite{ganea2021independent}. All models were evaluated using complex root mean square deviation (C-RMSD) and interface root mean square deviation (I-RMSD). C-RMSD is determined by superimposing the ground truth and predicted complex structures via the Kabsch algorithm \citep{kabsch1976solution} and computing the RMSD between all C-$\alpha$ coordinates. I-RMSD is determined by similarly aligning the interface residues of both complexes and computing the RMSD over interface C-$\alpha$ coordinates (within 8\AA{} of the binding partner). 

\section{Results}


Table ~\ref{tab:dips_test} reports the performances of the different methods on the DIPS test set. \ours{} achieves a C-RMSD median of 4.85, in line with \textsc{HDock} and significantly outperforming all other baselines. This performance is further confirmed when looking at the percentage of predictions below a specific threshold with \ours{} achieving respectively 42\% and 45\% of C-RMSD and I-RMSD below 2 \AA{}. 

\begin{table}[h!]
\caption{\textbf{Results on 100 samples from the DIPS test set.} The last three rows show our method's performance. The number of poses sampled from the diffusion model is in parentheses, and oracle refers to the setting where we can perfectly select the best pose out of the sampled ones. The methods highlighted with * do not use the same training data as our models and might be using parts of our test sets (e.g., to extract templates or as training examples) or have seen proteins more similar to our test set than the training set. Runtimes with $\dagger$ denote that the method can only run on CPU.}
\label{tab:dips_test}
\centering
\resizebox{\textwidth}{!}{%
\begin{tabular}{@{}lccccccccc@{}}
\toprule
\multicolumn{1}{c}{}  & \multicolumn{8}{c}{DIPS Test Set}                                                     & \multicolumn{1}{l}{} \\
 &
  \multicolumn{4}{c|}{Complex RMSD (\AA)} &
  \multicolumn{4}{c|}{Interface RMSD (\AA)} &
  Runtime (s) \\ \cmidrule(l){2-10} 
Methods &
  \%\textless{}2 &
  \%\textless{}5 &
  \multicolumn{1}{l}{\%\textless{}10} &
  \multicolumn{1}{c|}{Median} &
  \%\textless{}2 &
  \%\textless{}5 &
  \multicolumn{1}{l}{\%\textless{}10} &
  \multicolumn{1}{c|}{Median} &
  Mean \\ \midrule
\textsc{Attract}*               & 20 & 23 & 33 & \multicolumn{1}{c|}{17.17} & 20 & 22 & 38 & \multicolumn{1}{c|}{12.41} & 1285$^\dagger$                 \\
\textsc{HDock}*                 & \textbf{50} & \textbf{50} & 50 & \multicolumn{1}{c|}{6.23}  & \textbf{50} & 50 & 58 & \multicolumn{1}{c|}{\textbf{3.90}}  & 778$^\dagger$               \\
\textsc{ClusPro}*               & 12 & 27 & 35 & \multicolumn{1}{c|}{15.77} & 21 & 27 & 42 & \multicolumn{1}{c|}{12.54} & 10475$^\dagger$                 \\
\textsc{PatchDock}*             & 31 & 32 & 36 & \multicolumn{1}{c|}{15.25} & 32 & 32 & 42 & \multicolumn{1}{c|}{11.45} & 7378$^\dagger$                  \\
\textsc{AlphaFold-multimer}*              & 39  & 45  & 52 & \multicolumn{1}{c|}{8.61} & 45  & 47 & 58 & \multicolumn{1}{c|}{6.67} & 1560                    \\
\textsc{EquiDock}              & 0  & 8  & 29 & \multicolumn{1}{c|}{13.30} & 0  & 12 & 47 & \multicolumn{1}{c|}{10.19} & \textbf{3.88}                    \\
\ours{}(1)          & 34 & 41 & 46 & \multicolumn{1}{c|}{11.95} & 36 & 42 & 53 & \multicolumn{1}{c|}{8.60} & 4.2                      \\
\ours{}(40)          & 42 & \textbf{50} & \textbf{55} & \multicolumn{1}{c|}{\textbf{4.85}} & 45 & \textbf{52} & \textbf{63} & \multicolumn{1}{c|}{4.23}  & 153                     \\ \midrule
\ours{}(40) - oracle & 71 & 79 & 86 & \multicolumn{1}{c|}{0.67}  & 72 & 82 & 91 & \multicolumn{1}{c|}{0.54}  & 153                      \\ \bottomrule
\end{tabular}%
}
\end{table}

When limited to generating one sample for each complex, \ours{} still outperforms the majority of the baselines while having a significantly lower runtime than the search-based methods. Ensuring computational efficiency without a significant drop in performance is critical for computational screening applications like drug discovery and antibody design, where one needs to analyze a very large number of complexes.
 

We note that when we evaluate the model in such a way that we choose the complex with the smallest RMSD from the ones generated by the model, the performance exceeds that of all baselines by a large margin. As such, the performance of \ours{} can be significantly boosted by improving the used confidence model or designing more effective ranking methods. Additionally, this regime can be particularly beneficial for applications where it is desirable to have at least one high-quality recommendation among the predicted suggestions, e.g., if a practitioner is interested in discovering a single good structure based on prior knowledge from a set of proposals where the diversity of the proposals is important.


\begin{figure}[h!]
  \centering
   \includegraphics{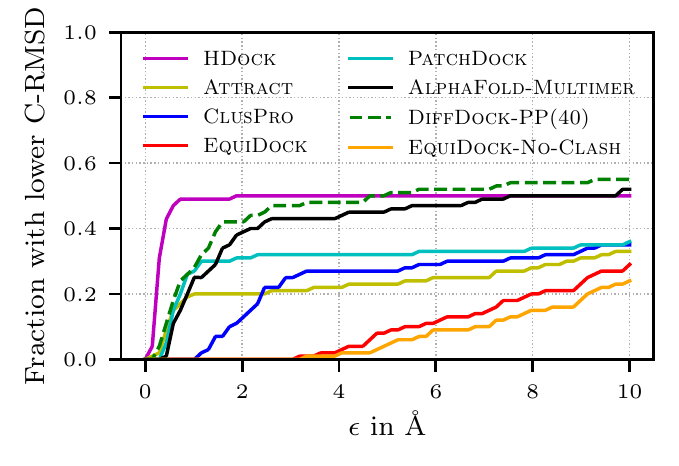}
\caption{Fraction of complexes with a C-RMSD $< \epsilon$ for different values of $\epsilon$ on 100 samples from the DIPS test set.}
\label{fig:cumulative}
\end{figure}

Figure \ref{fig:cumulative} shows, for all the considered methods, the fraction of predictions having a C-RMSD value below different thresholds ranging from 0 to 10. \ours{} outperforms most baselines for all threshold values, and outperforms \textsc{HDock} for thresholds higher than 5.

\begin{figure}[h!]
  \centering
   \includegraphics{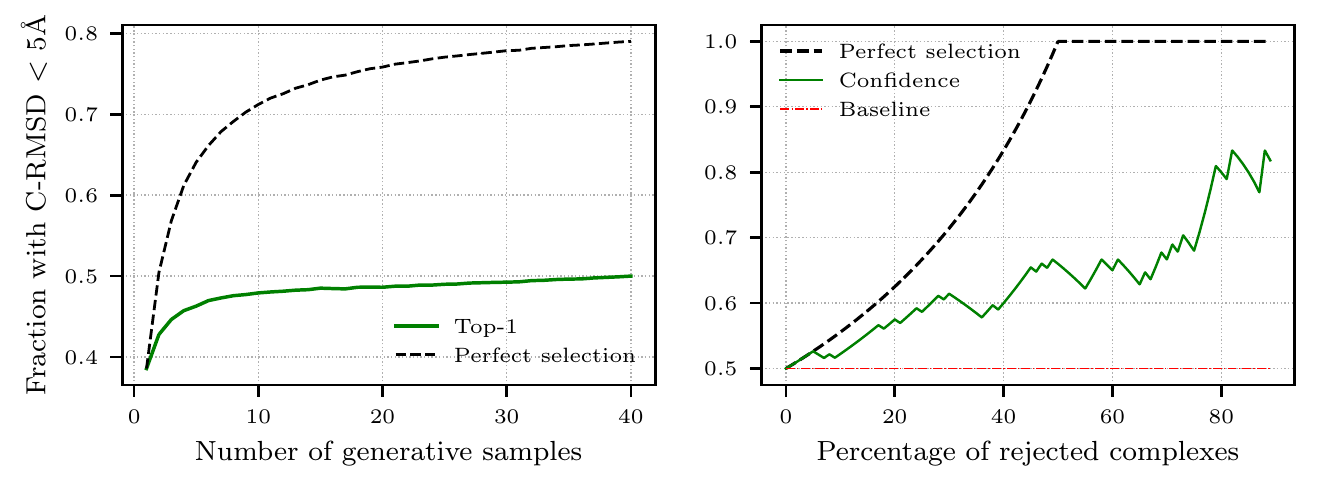}
\caption{\textbf{Left}: Performance of \textsc{DiffDock-PP} for increasing number of generative samples. “Perfect selection” is the theoretically best performance of the diffusion model assuming a perfectly accurate confidence model. \textbf{Right}: Fraction
of predictions with C-RMSD $<5$\AA{} considering only predictions for the part of the
dataset where \textsc{DiffDock-PP}(40) is most confident.}
\label{fig:confidence_calibration}
\end{figure}

To evaluate how good our confidence model is, we plot in Figure \ref{fig:confidence_calibration}-left the top-1 performance for different numbers of generated samples (this is done by picking the sample with the highest confidence score according to the confidence model) and compare it to the performance when we ignore the confidence score and instead pick the sample than minimizes the C-RMSD out of the generated samples. Performance here is defined as the fraction of samples having a C-RMSD below 5\AA{}. While the top-1 performance consistently increases with increasing number of generated samples, there still is a significant gap behind the performance in the perfect selection regime. This points out that \ours{} can achieve significantly better performance by improving the current confidence model or by developing a more involved selection algorithm.

The right plot of Figure \ref{fig:confidence_calibration} illustrates the selective accuracy of our model if we consider only complexes with confidence predictions above a certain threshold. We do this by ordering the final predicted complex structures by their assigned confidence prediction for all 100 complexes and removing the complexes with the lowest confidence score one by one. By removing predictions with lower confidence, we see an increase in the success rate of our prediction model.

We note that due to time constraints, it was not possible to properly evaluate our method on the Docking Becnhmark 5.5 (DB5.5) dataset \cite{vreven2015updates} and leave it as an interesting future extension to this work.






\section{Conclusion}

We present \textsc{DiffDock-PP}, a diffusion generative model for rigid protein-protein docking. Our approach is inspired by recent advancements in molecular docking \citep{corso2022diffdock}, which tackles docking via a generative model over ligand poses.
\textsc{DiffDock-PP} outperforms existing deep learning models and performs competitively against search-based methods at a fraction of their computational cost.
The effectiveness of our simple approach paves the way for further investigation into deep learning for modeling biomolecular interactions.

\section{Acknowledgements}

This material is based upon work supported by the National Science Foundation Graduate Research Fellowship under Grant No. 1745302. It is also supported by the Machine Learning for Pharmaceutical Discovery and Synthesis (MLPDS) consortium, the Abdul Latif Jameel Clinic for Machine Learning in Health, the DTRA Discovery of Medical Countermeasures Against New and Emerging (DOMANE) threats program, the DARPA Accelerated Molecular Discovery program and the Sanofi Computational Antibody Design grant. The authors thank Dr. Ricardo Acevedo Cabra and Prof. Dr. Massimo Fornasier for providing the opportunity to work on this project as part of the TUM Data Innovation Lab.




\bibliography{iclr2023_conference}
\bibliographystyle{iclr2023_conference}

\appendix
\section{Baselines: Experimental Details}
\label{appendixA}
In this section we give some details on how we evaluated the baselines to which we compared our method. 

For \textsc{HDock}, we used the HDOCKlite package that can be downloaded from \texttt{http://huanglab.phys.hust.edu.cn/software/hdocklite/}.

For \textsc{EquiDock}, we used the official implementation found in \texttt{https://github.com/octavian-ganea/equidock\_public}.

For \textsc{AlphaFold-Multimer}, we first extracted FASTA sequences from the complexes using the residue names given in their respective PDB files. Each complex yielded one FASTA file with two chains. Then, we ran AlphaFold-Multimer v2.3.0 using the official implementation of the inference pipeline provided in \texttt{https://github.com/deepmind/alphafold}. This implementation utilizes the following datasets: BFD, MGnify, PDB70, PDB (structures in the mmCIF format), PDB seqres, UniRef30 (formerly UniClust30), UniProt, and UniRef90.

For \textsc{Attract}, \textsc{ClusPro} and \textsc{PathDock}, we used the results reported by \cite{ganea2021independent}.


\end{document}